\documentclass[aps,twocolumn,showpacs,superscriptaddress]{revtex4}
\usepackage{graphicx}   
\usepackage{epstopdf}
\usepackage{epsfig}
\usepackage{dcolumn}
\usepackage{bm}	
\usepackage{amsmath}			
\usepackage{amsfonts}			
\usepackage{amssymb}			
\usepackage{latexsym}			
\usepackage{color}

\begin{document}

\title{Nonlinear dynamics of the ion Weibel-filamentation instability: an analytical model for the evolution of the plasma and spectral properties}
\author{C. Ruyer}\email{charles.ruyer@polytechnique.edu}
\affiliation{CEA, DAM, DIF, F-91297 Arpajon, France}

\author{L. Gremillet}\email{laurent.gremillet@cea.fr}
\affiliation{CEA, DAM, DIF, F-91297 Arpajon, France}

\author{A. Debayle}
\affiliation{CEA, DAM, DIF, F-91297 Arpajon, France}

\author{G. Bonnaud}
\affiliation{CEA, Saclay, INSTN, F-91191 Gif-sur-Yvette, France}

\begin{abstract}
We present a predictive model of the nonlinear phase of the Weibel instability induced by two symmetric, counter-streaming ion beams in the non-relativistic
regime.  This self-consistent model combines the quasilinear kinetic theory of Davidson \emph{et al.} [Phys. Fluids \textbf{15}, 317 (1972)] with a simple description
of current filament coalescence. It allows us to follow the evolution of the ion parameters up to a stage close to complete isotropization, and is thus of prime interest to
understand the dynamics of collisionless shock formation. Its predictions are supported by 2-D and 3-D particle-in-cell simulations of the ion Weibel instability.
The derived approximate analytical solutions reveal the various dependencies of the ion relaxation to isotropy. In particular, it is found that the influence of
the electron screening can affect the results of simulations using an unphysical electron mass.

\end{abstract}

\maketitle


First-principles kinetic simulations of plasma collisions governed by electromagnetic effects are now made possible using massively parallel particle-in-cell (PIC) codes,
hence paving the way to quantitative modeling of a number of high-energy astrophysical scenarios \cite[]{Kato_2008,Spitkovsky_2008}. The turbulent shocks possibly
arising from plasma instabilities in these systems are believed to be responsible for the generation of nonthermal particles and radiation \cite[]{Drury_1983,Malkov_2001,Piran2004}.
In this context, many simulation studies have demonstrated the ability of the Weibel-filamentation instability \cite[]{Weibel_1959, Fried_59, ppr_sagdeev_66, Medvedev1999, Achterberg_2007,Achterberg_2007b} to provide the electromagnetic turbulence required for efficient dissipation of the flow energy and Fermi-type acceleration processes
\cite[]{Spitkovsky_2008,Spitkovsky_2008b,Martins2009}. These numerical advances go along with experimental progress towards the laser-driven generation of collisionless
turbulent shocks in the laboratory \cite[]{Kuramitsu_2011,POP_Kugland_2013,Fox_2013,Huntington_2013}.

Collisionless shocks developing in electron-ion plasmas may be of laminar or turbulent nature depending on the type (electrostatic or electromagnetic) of the dominant
underlying instability \cite[]{Stockem_SR_2014}. In this work, we concentrate on initially unmagnetized electron-ion systems whose collective dynamics is eventually
ruled by the electromagnetic ion Weibel instability, which may evolve into a turbulent shock. While this problem has inspired a number of numerical studies
\cite[]{Kato_2008,Spitkovsky_2008,Martins2009,Fox_2013, Huntington_2013,Stockem_SR_2014,Stockem_PRL_2014}, there is as yet no analytical model of the nonlinear
evolution of the ion Weibel instability leading to shock formation. Our goal is to provide such a description within the simplifying assumption of homogeneous and infinite
colliding plasmas of equal densities and temperatures. Our paper is organized as follows. In Sec. \ref{sec:transition}, we first analyze the results of a reference
PIC simulation, pointing out the transition from the early-time electron-driven phase, associated with various fast-growing modes, to the ion-driven phase ruled by the ion
Weibel instability.  In Sec. \ref{sec:ql_model}, we present a set of quasilinear equations describing the evolution of the ion parameters in the time-varying magnetic turbulence
generated by the Weibel instability \cite[]{Davidson_1972}. Approximate relations between the plasma and spectral parameters are obtained and successfully confronted to
a number of PIC simulations.  Our model is then made self-consistent by the inclusion of  a simple description of current filament coalescence. The derived analytical solutions
are shown in good agreement with PIC simulation results.  Our concluding remarks are given in Sec. \ref{sec:conclusions}.

\section{Transition between electron and ion instability regimes}\label{sec:transition}

\begin{figure*}[t!]
\centerline{\includegraphics[scale=1]{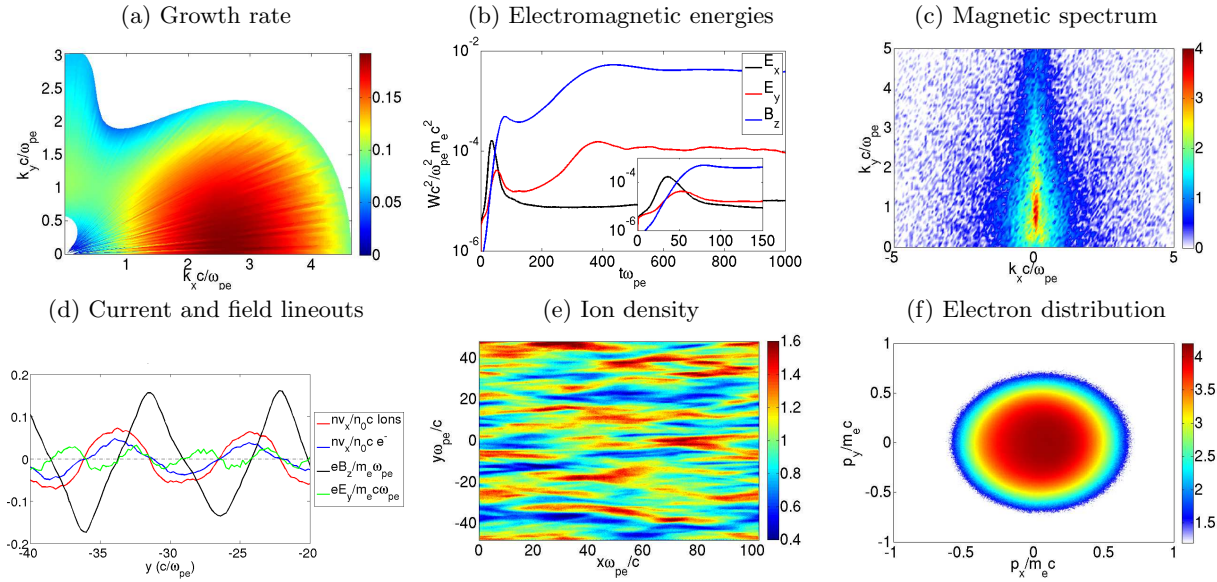}}

\caption{\label{fig:filam_ion} 
(a) Growth rate $\Gamma/\omega_{pe}$ in the ($k_x, k_y$)-plane for the plasma parameters $m_i=100m_e$, $v_e=v_i=\pm 0.2c$
and $T_e=T_i=m_ec^2/100$.  The real frequency vanishes for all the illustrated unstable modes. (b) Temporal evolution of the space-integrated $E_x$ (black),
$E_y$ (red) and $B_z$ (blue) electromagnetic energies. (c) Total ion density (normalized to the total initial density $n_i^{(0)}$) at  $\omega_{pe}t=690$. 
(d) $p_x-p_y$ electron phase space at  $\omega_{pe}t=690$.
(e) Magnetic spectrum, $e \vert B_z(k_x,k_y)\vert /m_e\omega_{pe}$ (in $\log_{10}$ scale), at  $\omega_{pe}t=690$.
(f) Transverse lineouts of the $E_y$ and $B_z$ fields and of the particles' mean $x$-velocity at  $\omega_{pe}t=690$. 
}
\end{figure*}

The PIC simulations presented in this study have been performed using the parallel code \textsc{calder} \cite[]{Lefebvre_2003}. All of them consider plasma systems made
of two symmetric, counter-streaming electron-ion beams  of uniform density and velocity profiles. Both the electrons and ions initially obey Maxwell-J\"uttner distributions of
temperatures $T_e =T_i$ and drift velocities $\mathbf{v_e}=\mathbf{v_i} = \pm v_0 \mathbf{\hat{x}}$. Periodic boundary conditions are used for the fields and particles in all directions.
Our reference two-dimensional (2-D) simulation, examined below, is characterized  by $T_{e,i}/m_ec^2=0.01$ and $v_0/c = 0.2$. We use singly-charged ions with a
reduced mass $m_i/m_e =100$ to alleviate the computational cost.  The density of each beam is normalized to the total electron density ($n_{e,i}=1/2$). The mesh sizes
are $\Delta x = \Delta y = 0.1c/\omega_{pe}$, where $\omega_{pe}$ is the total electron plasma frequency. The time step is $\Delta t= 0.069\omega_{pe}^{-1}$. The domain
size is $1024\Delta x \times 1024\Delta y$, that is, $102.4c/\omega_{pe} \times 102.4 c/\omega_{pe}$. Each cell initially contains 50 macro-particles per species, yielding
a total number  of about $2\times 10^8$ macro-particles. Third-order weight factors are employed along with Esirkepov's current deposition scheme \cite[]{Esirkepov_2001}.

In the relativistic, cold electron-positron systems considered in Refs. \cite[]{Bret_POP_2013,POP_Bret_2014}, the Weibel filamentation was found to prevail from
early on. By contrast, the non- or weakly-relativistic, warm electron-ion systems treated henceforth usually experience a progressive transition from an early phase
ruled by electron-driven modes to a regime ruled by the slower ion Weibel instability. Our reference simulation illustrates this transition. Figure \ref{fig:filam_ion}(a)
displays the theoretical growth rate map in the $(k_x,k_y)$ plane, computed from the electromagnetic dispersion relation (see the Appendix) with the initial plasma parameters.
The largest growth rate, $\Gamma_\mathrm{max}/\omega_{pe}\simeq 0.18$, corresponds to a longitudinal electrostatic mode located at ($k_x,k_y) \simeq (2.5,0)\omega_{pe}/c$.
The electromagnetic Weibel modes are located around the $k_y$-axis, with a maximum growth rate $\Gamma/\omega_{pe} \simeq 0.1$   
reached at the purely
transverse wave vector $(k_x,k_y)\simeq (0,1)\omega_{pe}/c$ \mbox{[Fig. \ref{fig:filam_ion}(a)]}. 
Note that the growth rate map's fastest-growing mode associated with a given wave vector. 
In the present case, these dominant modes are essentially driven by the electrons and, owing to the symmetry of the system, present a vanishing real
frequency. 
We verified that the calculation performed with immobile ions give negligible differences in the growth rates.

The predicted electrostatic character of the early interaction phase is confirmed by the time history of the integrated electromagnetic energies [\mbox{Fig. \ref{fig:filam_ion}(b)}]. 
At early times ($\omega_{pe}t \lesssim 40$), the $E_x$ energy is the dominant one, growing exponentially at an effective rate $\Gamma/\omega_{pe} \simeq 0.14$ consistent
with linear theory.  The associated longitudinal instability saturates at $\omega_{pe}t \simeq 40$ 
and subsequently decays away [\mbox{Fig. \ref{fig:filam_ion}(b)}]. 
The magnetic $B_z$ energy
then takes over, growing at a rate $\Gamma/\omega_{pe}\simeq 0.08$ close to the fastest Weibel mode of \mbox{Fig. \ref{fig:filam_ion}(a)}. A first magnetic saturation occurs
at $\omega_{pe}t \simeq 80$, which marks the end of the electron-governed Weibel instability. The instability then switches to an ion-driven regime \cite[]{Achterberg_2007,Achterberg_2007b,Nitin_2012},
initially characterized by exponentially increasing $B_z$ and $E_y$ energies (while the $E_x$ energy keeps stagnating at a low level). This growing phase comes to an
end at $\omega_{pe}t \simeq 500$, at which time the instability enters its nonlinear saturation phase, further studied in the following. This phase exhibits stagnating field
energies but also, as analyzed below, continuously evolving plasma and spectral distributions.

The transverse character of the magnetic instability is  evidenced in \mbox{Fig. \ref{fig:filam_ion}(c)}, which displays (in $\log_{10}$ scale) the magnetic spectrum
$\vert B_z(k_x,k_y)\vert$ at $\omega_{pe}t = 690$. The spectral energy is concentrated along the $k_y$-axis, peaking around $(k_x,k_y) \simeq (0,0.7)\omega_{pe}/c$.
The dominant ion contribution to the instability is demonstrated by the transverse field and current lineouts plotted in \mbox{Fig. \ref{fig:filam_ion}(d)}. The total ion
current appears more strongly modulated  than the electron current, so that the magnetic fluctuations are mainly induced by the ions. This feature, shown here at
$\omega_{pe}t=690$, is found to hold from $\omega_{pe}t \simeq 300$ onwards. As expected in its nonlinear regime, the ion Weibel instability generates, besides
current modulations, significant ion density fluctuations in the transverse direction [\mbox{Fig. \ref{fig:filam_ion}(e)}]. These ion density filaments also present
longer-wavelength longitudinal modulations that can be ascribed to magnetic coalescence processes \cite[]{Medvedev_2005, Achterberg_2007b}. The latter will
be shown to govern the nonlinear evolution of the magnetic turbulence. The partial electron screening of the ion density filaments accounts for the correlated growth
of the $E_y$ and $B_z$ energies seen in \mbox{Fig. \ref{fig:filam_ion}((b)}. From the lineouts of \mbox{Fig. \ref{fig:filam_ion}(d)}, the $E_y$ fluctuations have a typical
wavelength twice smaller than the $B_z$ fluctuations, which suggests that an approximate balance between the transverse electric and magnetic forces is established
on the electrons \cite[]{Dieckmann_POP_2009}. From this reasoning, the transverse electric field is expected to scale as $E_y \sim \frac{e}{2m_e} \partial_y A_x^2$,
where $A_x$ is the vector potential. Assuming $A_x \sim B_0 \cos (k_y y)/k_y$ (where $B_0$ is the magnetic field amplitude and $k_y$ the dominant transverse
wave number), one predicts the transverse electric field amplitude $E_0 \sim \frac{e}{2m_ek_y}B_0^2$. Using $eB_0/m_e\omega_{pe} =  0.15$ and $k_yc/\omega_{pe} = 0.7$,
one finds $eE_0/m_ec\omega_{pe} =  0.016$, in fair agreement with \mbox{Fig. \ref{fig:filam_ion}(d)}.

Due to the ensemble of electron-driven instabilities developing at early times [\mbox{Fig. \ref{fig:filam_ion}(a)}], the electrons turn out to be essentially isotropized
in the ion Weibel regime, as shown by their $p_x-p_y$ phase space at $\omega_{pe}t=690$ [\mbox{Fig. \ref{fig:filam_ion}(f)}]. 
More quantitatively, the average
$x$-velocity of each electron beam has dropped from $\vert v_e \vert =0.2c$ to $\vert v_e \vert \simeq0.06c$ at the beginning of the nonlinear ion Weibel phase
($\omega_{pe}t \simeq 500$),  while its temperature has increased to $T_e/m_ec^2 \simeq 0.04$ [see Figs. \ref{fig:vi02_ve02_m100}(d,f)]. 

\section{Quasilinear model of the ion Weibel filamentation}\label{sec:ql_model}

\subsection{Dominant unstable mode} \label{subsec:dominant_mode}

A major assumption of our model is that the main properties of the magnetic spectrum can be related to the instantaneous plasma parameters. In order to derive such
a relation, let us first examine the linear characteristics of the purely transverse ion Weibel instability. Its general dispersion relation follows from taking $\theta=\pi/2$
in Eq. \eqref{eq:dispe_2d}:
\begin{equation}
  (\omega^2 \varepsilon_{xx}-k_y^2c^2)\omega^2 \varepsilon_{yy} - \omega^4 \varepsilon_{xy}^2 = 0\, .
\end{equation}
In the case of symmetric counter-streaming plasmas, the off-diagonal tensor element vanishes, yielding the well-known simplified dispersion relation \cite[]{Weibel_1959}
\begin{equation}
  \omega^2 \varepsilon_{xx}-k_y^2c^2 = 0\, .
\end{equation}
For the sake of simplicity, we will restrict our analysis to non-relativistic systems initially described by two-temperature, drifting Maxwellians of the form
\begin{equation}\label{eq:bm}
f_s^{(0)}(\mathbf{v}) = \frac{m_s}{2\pi\sqrt{T_{sx}T_{sy}}} \exp{\left[-\frac{m_s(v_x - v_s)^2}{2T_{sx}}-\frac{m_s v_y^2}{2T_{sy}} \right]} \, ,
\end{equation}
where $m_s$ is the mass, $v_s$ is the drift velocity and $T_{xs}$ and $T_{sx}$ are, respectively, the longitudinal and transverse temperatures of the $s$th plasma species.
Using the expressions \eqref{eq:xibm1}, \eqref{eq:xibm2} and \eqref{eq:xibm3}, and exploiting the purely imaginary character of the Weibel modes ($\omega=i\Gamma$
with $\Gamma >0$), the dispersion relation can be recast as \cite[]{Davidson_1972}
\begin{equation}
  k_y^2c^2 +\Gamma^2 + \sum_s\omega_{ps}^2 -\sum_s \omega_{ps}^2(a_{s}+1) \Re \left[ 1+  \xi_s\mathcal{Z}\left( \xi_s\right) \right]=0  \, . \label{eq:dispew1}
\end{equation}
In the above equation,  $\mathcal{Z}$ denotes the plasma dispersion function \cite[]{Fried_Gell-Mann_1960}, $\omega_{ps}$ the plasma frequency of the $s$th species, 
$\xi_s = i\sqrt{m_s/2T_{ys}}\Gamma/k_y$,  $\Re$ the real part and  $a_s$ the anisotropy ratio of the $s$th species, defined as
\begin{equation}
  a_s = \frac{m_sv_{s}^2 +T_{sx}}{T_{sy}}-1 \label{eq:an}\, .
\end{equation}
We will now assume that, independently of their initial distribution, the electrons are almost completely isotropized ($a_e\sim 1$) in the nonlinear ion Weibel phase. 
Furthermore, they will be assumed hot enough so that $\vert \xi_e \vert \ll 1$, allowing us to use the small-argument expansion of $\mathcal{Z}$:
\begin{equation}\label{eq:tdz}
  \mathcal{Z}(\xi) = i\sqrt{\pi}\frac{k}{\vert k \vert} \exp{(-\xi^2)} - 2\xi  +O(\xi^3)\, .
\end{equation}
To leading order, we obtain $\vert \xi_e\mathcal{Z}(\xi_e) \vert  \simeq  \vert \sqrt{\pi}\xi_e \vert \ll 1 $. For instance, solving \mbox{Eq. \eqref{eq:dispew1}} with the
plasma parameters measured in the reference simulation at $\omega_{pe}t=500$ ($T_{ey} \simeq T_{iy} \simeq 0.04m_ec^2$ 
leads to $\vert \xi_e \vert \simeq 0.1$
and $ \vert \xi_i\mathcal{Z} \vert \simeq 0.15$. This term will thus be neglected in the bracketed factor of Eq. \eqref{eq:dispew1}.

The ion response will be assumed to fulfill $\vert \xi_i \vert \lesssim 1$, so that we retain the leading term of $\xi_i\mathcal{Z}(\xi_i) \simeq  \sqrt{\pi}\xi_i$ in
\mbox{Eq. \eqref{eq:dispew1}}. There results the approximate dispersion relation 
\begin{equation}\label{eq:dispew2}
  \omega_{pi}^2(a_{i} +1 ) \sqrt{\frac{\pi m_i}{2T_{iy}}} \Gamma + k_y^2 \left(k_y^2 - \frac{\omega_{pe}^2a_e}{c^2} -\frac{\omega_{pi}^2 a_i}{c^2} \right)  = 0 \, .
\end{equation}
The growth rate is readily solved as
\begin{equation}\label{eq:wg2}
\Gamma \simeq \sqrt{\frac{2T_{iy}}{\pi m_i}} \vert k_y \vert  \frac{k_\mathrm{max}^2c^2 - k_y^2c^2}{\omega_{pi}^2(a_i +1 ) }  \, ,
\end{equation}
where $k_\mathrm{max}$ denotes the upper bound of the Weibel-unstable domain:
\begin{equation}
k_\mathrm{max} = c^{-1} \sqrt{\omega_{pe}^2a_e + \omega_{pi}^2a_i} \, . \label{eq:kmax} 
\end{equation}
Equation \eqref{eq:wg2} is formally similar to that derived for the electron Weibel instability in the weak growth rate limit \cite[]{Davidson_1972}.
The only difference lies here in $k_\mathrm{max}$, which involves both the ion and the electron anisotropies. 

\begin{figure}
\centerline{\includegraphics[scale=1]{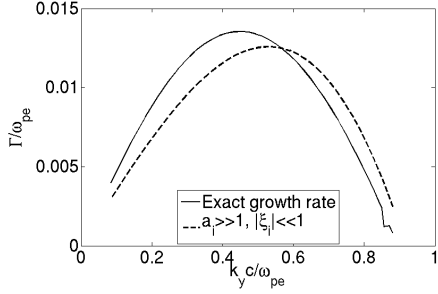}}
\caption{\label{fig:gamma_approx2} $k$-dependence of the Weibel growth rate for the plasma parameters $v_i=\pm0.2c$, $v_{e}=0$, $T_{ex,y}=T_{ix,y}=0.05m_ec^2$
and $m_i/m_e=25$: comparison between the exact solution of Eq. \eqref{eq:dispew1} (solid line) and the estimate, Eq. \eqref{eq:wg2}.
}
\end{figure}

The accuracy  of the estimate, Eq. \eqref{eq:wg2}, is illustrated in Fig. \ref{fig:gamma_approx2} for the plasma parameters $v_i=\pm0.2c$, $v_e=0$, 
$T_{ex,y}=T_{ix,y}=0.05m_ec^2$ and $m_i=25m_e$. Correct agreement is found between the exact and approximate growth rate curves, with respect
to both their general shapes and the location and amplitude of the dominant mode.  

In the following, we will assume that most of the magnetic energy is contained in the spectral region surrounding the fastest-growing wave number $k_\mathrm{sat}$.
Making use of Eq. \eqref{eq:wg2}, the solution of $\partial_{k_y} \Gamma = 0$ is
\begin{equation}\label{eq:ksat1}
  k_\mathrm{sat}  \simeq \frac{k_\mathrm{max}}{\sqrt{3}}  \, .
\end{equation}
To further simplify our analysis, we will henceforth use the approximation $k_\mathrm{sat}\simeq 0.5k_\mathrm{max}$. Moreover, we will take the
large-ion-anisotropy limit $a_i \gg1$, consistently with our focus on the isotropization process of initially highly anisotropic ion populations.
The $a_i \gg1$ limit will be assumed to hold (at least marginally) up to a stage close to ion isotropization ($a_i \gtrsim 2$). In an inhomogeneous
system, the latter stage should approach shock formation. There follow the estimates
\begin{align}
  &k_\mathrm{max} \simeq \omega_{pi}\sqrt{a_i}/c \, , \\
  &\Gamma_{k_y}  \simeq \sqrt{\frac{2T_{iy}}{\pi m_i}}  \left(1-\frac{k_y^2}{k_\mathrm{max}^2} \right) k_y\, ,  \label{eq:gamma1_5} \\
  &\xi_i  \simeq i \frac{1}{\sqrt{\pi}}\left(1-\frac{k_y^2}{k_\mathrm{max}^2}\right) \, . \label{eq:xii}
\end{align} 
Therefore, for $a_i \gg 1$, it is found that $\xi_i$ depends only on the ratio $(k_y/k_\mathrm{max})^2$. Anticipating on the next section, we define
$\xi_\mathrm{sat} \equiv \xi_i(k_y=k_\mathrm{sat}) \simeq 0.4i$. The small-argument expansion of  $\mathcal{Z}(\xi_i)$ in Eq. \eqref{eq:dispew1} is
then marginally valid. 

\subsection{Temporal evolution of the plasma parameters}\label{sec:tepp}

\subsubsection{Quasilinear equations}

The above formulae will serve to relate to the dominant wave vector, $k_\mathrm{sat}$, to the ion anisotropy ratio deduced from the spatially averaged
ion distribution functions, $\langle f_s \rangle(\mathbf{v},t)$. The evolution of the latter due to non-resonant wave-particle interaction in the Weibel magnetic
turbulence will be described in the framework of quasilinear kinetic theory
\cite[]{Montes_JPP_1970,Davidson_1972, Sadovskii_2001,Pokhotelov_2011,POP_Hellinger_Passot_2013}:
\begin{align}
 &\partial_t \langle f_s \rangle (\mathbf{v},t)=  -i \sum_{k_y} \frac{\omega_{pi}^2\vert B_{k_y} \vert^2}{\mu_0n_sm_sc^2k_y^2} \nonumber \\
 &\times \left[ -k_yv_x\partial_{v_y} + (i\Gamma_{-k_y}+k_yv_y)\partial_{v_x} \right] \nonumber \\
 &\times \left[\frac{ k_yv_x\partial_{v_y} +(i\Gamma_{k_y}-k_yv_y)\partial_{v_x} }{i\Gamma_{k_y}-k_yv_y} \right] \langle f_s \rangle (\mathbf{v},t) \, ,\label{eq:ql_dav} 
\end{align} 
where $\mu_0$ is the magnetic permittivity of vacuum.  We recall that the quasilinear kinetic theory for the Weibel instability is valid provided
$\vert \xi_s \vert \lesssim 1$, as already assumed in the previous section. The ability of the quasilinear theory to capture the nonlinear evolution
of the Weibel instability has been demonstrated by Davidson \emph{et al.} \cite[]{Davidson_1972}.

Assuming that the ion distribution functions remain of the bi-Maxwellian form, Eq. \eqref{eq:bm}, the three first moments of Eq. \eqref{eq:ql_dav} give
a set of differential equations on the mean ion drift velocities and temperatures:
\begin{align}
 &n_s\partial_t v_{s}  = -\sum_{k_y} \frac{\omega_{ps}^2}{k_y^2c^2} \frac{v_{s}}{T_{sy}} \Re\left[  1+\xi_s\mathcal{Z}(\xi_s) \right] \frac{\partial_t \vert B_{k_y} \vert^2}{\mu_0} \, , \label{eq:qlv}\\ 
 &n_s\partial_t T_{sy}  = \sum_{k_y} \frac{\omega_{ps}^2}{k_y^2c^2} (a_{s}+1) \Re\left[  1+\xi_s\mathcal{Z}(\xi_s) \right]\frac{\partial_t \vert B_{k_y} \vert^2}{\mu_0} \, , \label{eq:qlty}\\
&n_s\partial_t K_{sx}  = -\sum_{k_y} \frac{\omega_{ps}^2}{k_y^2c^2}  \Re\left[2(a_{s}+1)(1+\xi_s\mathcal{Z}(\xi_s)) -1\right] \nonumber \\
 & \times \frac{\partial_t \vert B_{k_y} \vert^2}{\mu_0}\, ,\label{eq:qltx}
\end{align}
with $K_{sx} = T_{sx} + m_sv_{s}^2$, the $x$-momentum flux. We have also exploited the relation
\begin{equation}\label{eq:partialb}
  \partial_t \vert B_{k_y} \vert^2 = 2 \Gamma_{k_y}  \vert B_{k_y} \vert^2 \, .
\end{equation}

\subsubsection{Approximate solutions of the quasilinear equations} \label{subsubsec:qlres}

In order to make analytical progress, we will make use of the approximation 
\begin{align}
  \sum_{k_y} \frac{\omega_{pi}^2}{c^2} \Re\left[1 + \xi_{k_y}\mathcal{Z}( \xi_{k_y} )\right]\frac{\partial_t \vert B_{k_y}\vert^2}{\mu_0k_y^2} \nonumber \\
  \simeq \frac{n_iZ_i^2}{m_i}  \Re\left[1 + \xi_\mathrm{sat}\mathcal{Z}( \xi_\mathrm{sat} )\right] S_p \, , \label{eq:approximation}
\end{align}
where $Z_i$ is the ion charge number and the spectral parameter $S_p$ is defined as
\begin{align} \label{eq:sp}
  S_p = e^2\sum_{k_y}   \frac{\vert B_{k_y}\vert^2}{k_y^2 }  =e^2 \sum_{k_y}\vert A_{k_y}\vert^2\, ,
\end{align}
which is homogeneous to the square of a momentum and where the sum runs over the positive and negatve wavevectors. 
Within the range of validity of \mbox{Eq. \eqref{eq:xii}}, the factor
$\alpha_i \equiv 1 + \xi_\mathrm{sat}\mathcal{Z}( \xi_\mathrm{sat})$ is a constant ($\alpha_i \simeq 0.5$), independent of the
plasma parameters. Equations  \eqref{eq:qlv}, \eqref{eq:qlty} and \eqref{eq:qltx} can then be recast as
\begin{align}
  &\partial_t v_i  = -\frac{Z_i^2}{m_i}  \alpha_i  \frac{v_i}{T_{iy}}  \partial_t S_p\, , \label{eq:qlvi} \\
  &\partial_t T_{iy}  = \frac{Z_i^2}{m_i}  \alpha_i (a_i+1)  \partial_t  S_p\, , \label{eq:qltyi} \\
  &\partial_t K_{ix}  = -\frac{Z_i^2}{m_i}   \left(2 \alpha_ia_i + 2\alpha_i -1\right) \partial_t S_p \, . \label{eq:qltxi} 
\end{align}

\begin{figure}
\centerline{\includegraphics[scale=1]{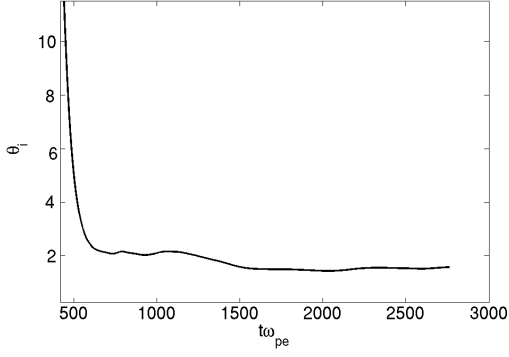}}
\caption{\label{fig:thetai} Temporal evolution of the quantity $\theta_i$ defined by Eq. \eqref{eq:thetai} in the reference simulation 
($v_i=v_e=\pm0.2c$ and $m_i=100m_e$).}
\end{figure}

An additional simplification can be made upon realizing that the ratio of Eqs. \eqref{eq:qltyi} and \eqref{eq:qltxi} is essentially constant in
the $a_{i}\gg1$ limit:
\begin{align}
  \frac{\partial_t K_{ix}}{\partial_t T_{iy}}\equiv - \theta_i &= -\frac{2\alpha_ia_{i}+2\alpha_i-1}{\alpha_i(a_{i}+1)} \, ,\label{eq:thetai} \\
  &\simeq -2\left[1  + \frac{1}{a_{i}}\left(1- \frac{1}{\alpha_i} \right)   \right] \, .
\end{align}
The validity of this approximation is supported by \mbox{Fig. \ref{fig:thetai}}, which plots the time history of $\theta_i$ from  the reference
simulation.  This quantity  is seen to rapidly drop down to a saturated value $\sim2$ once the ion Weibel instability enters its nonlinear
phase ($\omega_{pe}t \gtrsim 500$). At later times, $\theta_i$ slowly decreases (reaching $\simeq 1.8$ at $\omega_{pe}t = 2700$) as a result of
decreasing ion anisotropy [see Fig. \ref{fig:vi02_ve02_m100}(e)]. We will then  assume that $\theta_i$ is a quasi-constant (to leading order in
$1/a_i$) and neglect its time derivatives. Equation \eqref{eq:thetai} can then be readily integrated, giving
\begin{equation}\label{eq:thteta_i}
  K_{ix} = K_{ix}^{(0)} -\theta_i \left(T_{iy} - T_{iy}^{(0)} \right)\, ,
\end{equation}
where the notation $X^{(0)}$ stands for $X(t=0)$. Plugging Eq. \eqref{eq:thteta_i} into \eqref{eq:qltyi} leads to
\begin{equation}
  \frac{T_{iy}}{K_{ix}^{(0)} -\theta_i(T_{iy} - T_{iy}^{(0)})} \partial_t T_{iy}  = \frac{Z_{i}^2}{m_i} \alpha_i \partial_t  S_p\, . \label{eq:qltyi1}
\end{equation}
For practical reasons, we define the parameter 
\begin{equation}
  K_{\theta_i} = \theta_i T_{iy}^{(0)} + T_{ix}^{(0)} + m_i v_i^{(0)2}\, , \label{eq:ratioethetai}
\end{equation}
which has the dimension of an energy, and where $v_i^{(0)}\equiv v_0$.
For the typical value $\theta_i = 2$, we have $K_2 = 2T_{iy}^{(0)} + T_{ix}^{(0)} + m_iv_i^{(0)2}$.
The integration of Eq. \eqref{eq:qltyi1} is straightforward, yielding
\begin{equation}
  T_{iy} - T_{iy}^{(0)} +\frac{K_{2}}{\theta_i}\ln\left(\frac{K_{2} -\theta_i T_{iy}}{K_{2} -\theta_i T_{iy}^{(0)}} \right) =
  -\theta_i\frac{Z_i^2}{m_i} \alpha_i (S_p - S_p^{(0)}) \, . \label{eq:qltyi2}
\end{equation}
The Taylor expansion of the logarithmic term for $T_{iy}/K_{2}\le 1/a_i\ll 1$ (high-anisotropy limit) gives, to leading order, 
\begin{equation}
  T_{iy} \simeq \sqrt{T_{iy}^{(0)2}  + 2\frac{Z_i^2}{m_i} \alpha_iK_{2} \left(S_p - S_p^{(0)}\right)} \, . \label{eq:qltyi3}
\end{equation}
Inserting Eq. \eqref{eq:qltyi3} into \eqref{eq:qlvi} yields
\begin{equation}
  \partial_t v_i   \simeq -\frac{Z_i^2}{m_i} \alpha_i \frac{v_{i}\partial_t  S_p}{\sqrt{T_{iy}^{(0)2}  + 2\frac{\omega_{pi}^2}{n_ic^2} \alpha_iK_{2} \left(S_p - S_p^{(0)}\right)}}\, ,
\label{eq:qlvi1}
\end{equation}
which may be readily integrated as
\begin{equation}
v_i  \simeq v_0  \exp\left[ -2\frac{ T_{iy}(S_p)- T_{iy}^{(0)} }{ K_{2} }\right] \, ,\label{eq:qlvi2}
\end{equation}
where $T_{iy}$ verifies Eq. \eqref{eq:qltyi3}. Combining Eqs. \eqref{eq:thteta_i} and \eqref{eq:qltyi3} allows us to solve for $K_{ix}$:
\begin{equation}
 K_{ix} \simeq K_{2} -\theta_i  \sqrt{T_{iy}^{(0)2}  + 2\frac{Z_i^2}{m_i} \alpha_iK_{2} \left(S_p - S_p^{(0)}\right)} 
  \, . \label{eq:qlkix1}
\end{equation}
There follows the anisotropy ratio
\begin{equation}
 a_i \simeq \frac{K_{2} }{\sqrt{T_{iy}^{(0)2}  + 2\frac{Z_i^2}{m_i} \alpha_i K_{2}\left(S_p - S_p^{(0)}\right)}   }-2
  \, , \label{eq:qlani1}
\end{equation}
and the spectral parameter
\begin{equation}
 S_p - S_p^{(0)} \simeq \frac{m_i}{2Z_i^2\alpha_i K_{2}}\left[\left( \frac{K_{2}}{2+a_i}  \right)^2-T_{iy}^{(0)2}\right]
  \, . \label{eq:qlsp}
\end{equation}

To summarize, using non-resonant quasilinear theory \cite[]{Davidson_1972}, we have expressed the ion parameters $v_i$, $a_i$ $K_{ix}$ and
$T_{iy}$ in terms of the instantaneous spectral parameter $S_p$, independently of the time history of the ion Weibel-governed system. The
reader should be reminded that these approximate relations are valid in the limit of essentially  isotropic electrons and highly anisotropic
ions. 
\begin{figure*}[tbh!]
\centerline{\includegraphics[scale=1]{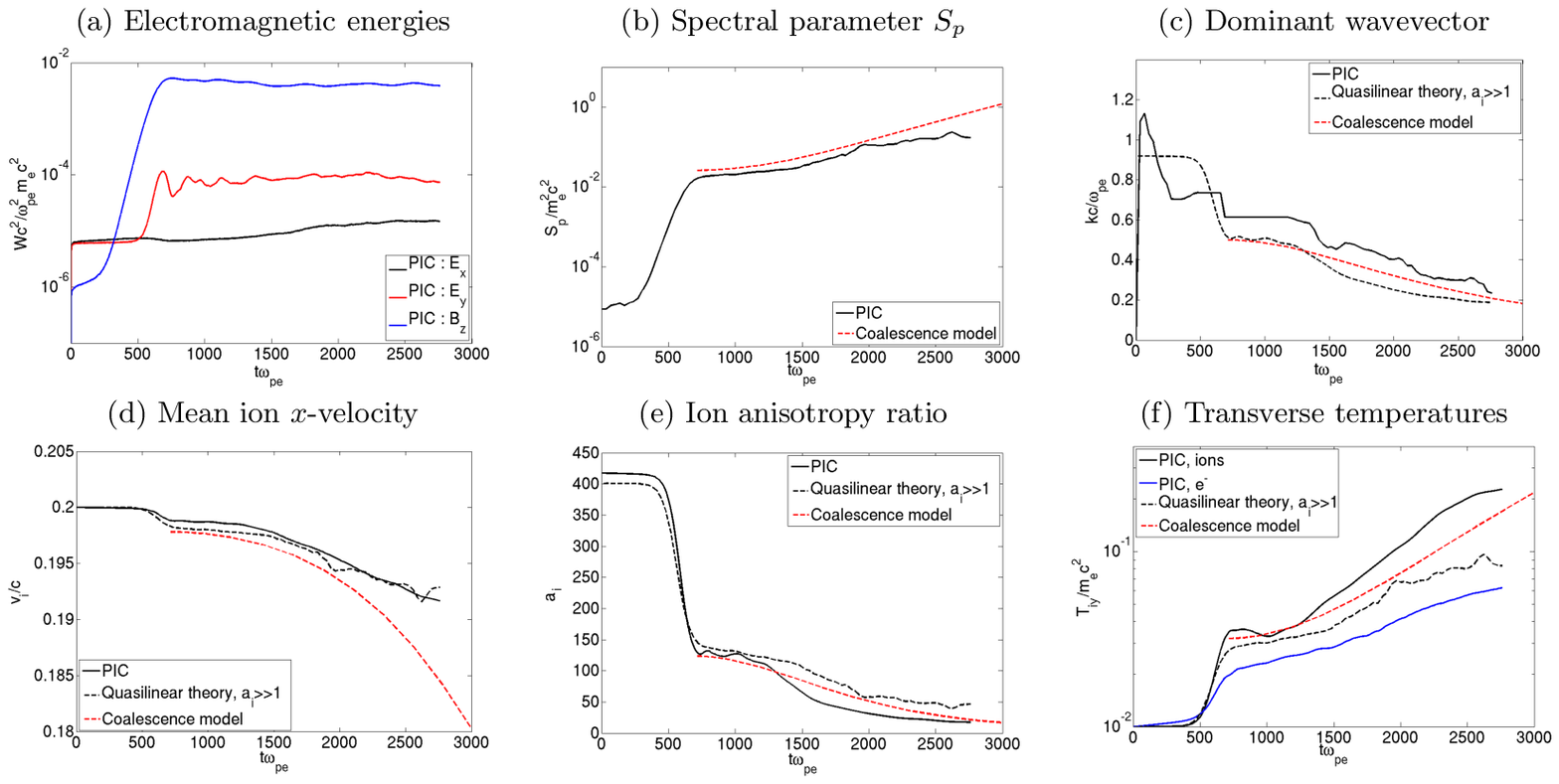}}
\caption{\label{fig:vi02_ve0_m100}
2-D simulation with initial conditions: $v_i =\pm0.2c$, $v_e=0$, $T_{e,i}/m_ec^2=0.01$, $m_i/m_e=100$.
(a) Temporal evolution of the mean electromagnetic energies (normalized to $m_ec^2\omega_{pe}^2/c^2$). 
(b) Temporal evolution of $S_p/(m_ec)^2$ from the simulation (black solid line).
(c) Temporal evolution of $k_\mathrm{sat}c/\omega_{pe}$ maximizing the $B_z$-spectrum from the simulation (black solid line) and
from Eq. \eqref{eq:ksat1} (black dashed line).
Temporal evolutions of (d) $v_i/c$, (e) $a_i$ and (f) $T_{i,ey}/m_ec^2$ from the simulation (black solid lines) and from Eqs. \eqref{eq:qlvi2}, \eqref{eq:qlani1}
and \eqref{eq:qltyi3}, respectively (black dashed lines), extracting $S_p$ from the simulation [black solid line of (a)]. 
The analytical predictions, Eqs. \eqref{eq:lambda_analytic}, \eqref{eq:anifinal}-\eqref{eq:vifinal}, are superposed as red dashed lines.
}
\end{figure*}

\subsection{Comparison with PIC simulation results}\label{subsec:pic}

This section gathers the PIC simulation results obtained using a variety of periodic (2-D or 3-D) geometries and plasma parameters, and confronts them to
the above analytical expressions. The latter will be computed using the PIC-predicted values of either the spectral parameter, $S_p(t)$, or the ion anisotropy
ratio, $a_i(t)$.

\subsubsection{2-D periodic simulations}\label{subsubsec:pic2d}

\begin{figure*}[tbh!]
\centerline{\includegraphics[scale=1]{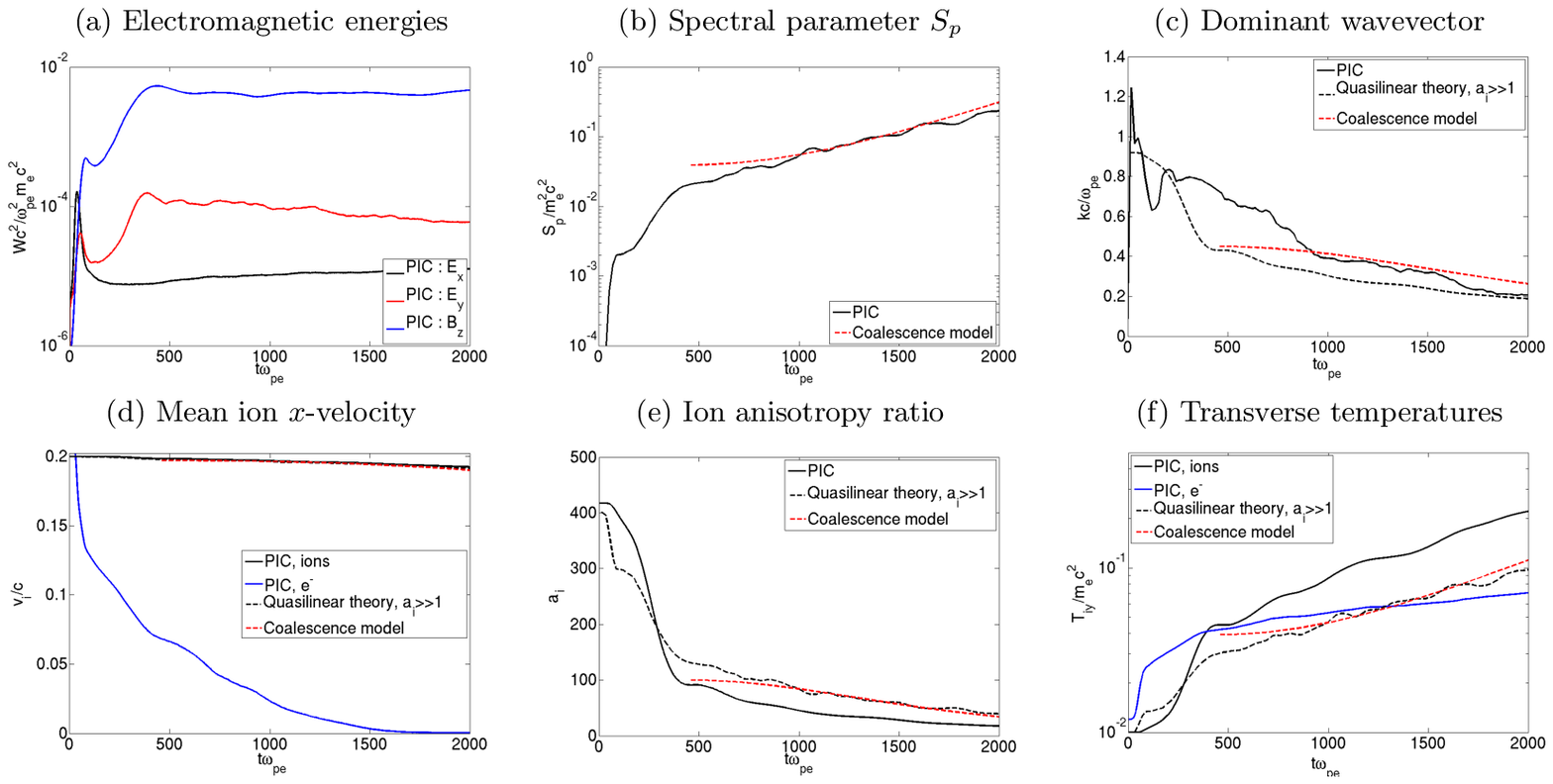}}
\caption{\label{fig:vi02_ve02_m100}
2-D simulation with initial conditions: $v_i =\pm0.2c$, $v_e=\pm 0.2c$, $T_{e,i}/m_ec^2=0.01$, $m_i/m_e=100$.
(a) Temporal evolution of the mean electromagnetic energies (normalized to $m_ec^2\omega_{pe}^2/c^2$). 
(b) Temporal evolution of $S_p/(m_ec)^2$ from the simulation (black solid line).
(c) Temporal evolution of $k_\mathrm{sat}c/\omega_{pe}$ maximizing the $B_z$-spectrum from the simulation (black solid line) and
from Eq. \eqref{eq:ksat1} (black dashed line).
Temporal evolutions of (d) $v_i/c$, (e) $a_i$ and (f) $T_{e,iy}/m_ec^2$ from the simulation (black solid lines) and from Eqs. \eqref{eq:qlvi2}, \eqref{eq:qlani1}
and \eqref{eq:qltyi3}, respectively (black dashed lines), extracting $S_p$ from the simulation [black solid line of (a)]. 
The analytical predictions, Eqs. \eqref{eq:lambda_analytic}, \eqref{eq:anifinal}-\eqref{eq:vifinal}, are superposed as red dashed lines.
}
\end{figure*}

For the 2-D PIC simulations, the spectral parameter $S_p$  is computed using a discrete Fourier transform in the transverse $y$ direction,
averaged along the $x$ direction: 
\begin{equation}
S_p^{2D} = e^2 \sum_{k_y} \frac{ \langle \vert DFT_y(B_z) \vert^2 \rangle_x  }{k_y^2}  \, . \label{eq:sp2d}
\end{equation}
Since we neglect the electron anisotropy in our quasilinear model, let us first consider a plasma system with initially isotropic
electrons. This configuration is exemplified in \mbox{Figs. \ref{fig:vi02_ve0_m100}(a-f)}, which gather the results of a simulation
run with $v_i/c=\pm0.2$, $v_e=0$, $T_{e,i}/m_ec^2=0.01$ and $m_i/m_e=100$ (black solid curves). The numerical resolution of the
corresponding dispersion relation (not shown) predicts that the fastest-growing mode ($\Gamma_\mathrm{max}/\omega_{pe} \simeq 0.013$)
is of the Weibel kind, and that there is no unstable longitudinal mode. These predictions are confirmed by \mbox{Fig. \ref{fig:vi02_ve0_m100}(a)},
which shows that the system's evolution is ruled from the start by the magnetic field growth. The $B_z$ energy exponentially increases
during the time interval $200 \lesssim \omega_{pe}t \lesssim 600$, at an effective rate $\Gamma/\omega_{pe} \simeq 0.01$, close to the
theoretical value. At the saturation time ($\omega_{pe}t�\simeq 600$), the $B_z$ energy is about 40 times larger than the $E_y$ energy. Later on,
although the field energies vary very weakly, the ion parameters ($v_i$, $a_i$, $T_{iy}$) continuously evolve [\mbox{Figs. \ref{fig:vi02_ve0_m100}(d-f)}].
These distinct behaviors point to a  time-changing magnetic spectrum, as confirmed by the increasing (resp. decreasing) trends of $S_p$
(resp. $k_\mathrm{sat}$) observed in \mbox{Figs. \ref{fig:vi02_ve0_m100}(b,c)}. 

\begin{figure*}[tbh!]
\centerline{\includegraphics[scale=1]{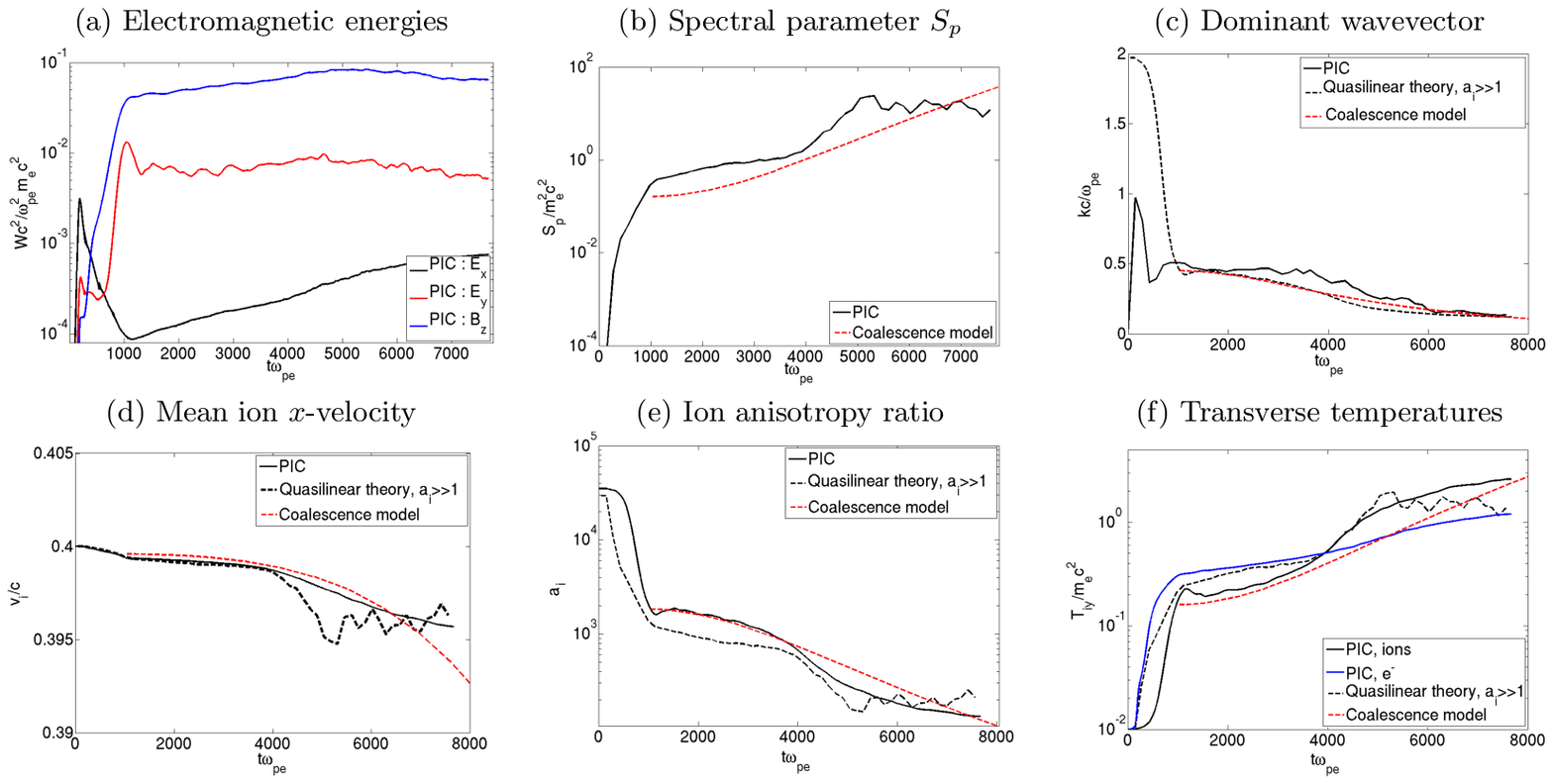}}
\caption{\label{fig:vi04_ve0_m1836}
2-D simulation with initial conditions: $v_i =\pm0.4c$, $v_e=0$, $T_{e,i}/m_ec^2=0.01$, $m_i/m_e=1836$
(a) Temporal evolution of the mean electromagnetic energies (normalized to $m_ec^2\omega_{pe}^2/c^2$). 
(b) Temporal evolution of $S_p/(m_ec)^2$ from the simulation (black solid line).
(c) Temporal evolution of $k_\mathrm{sat}c/\omega_{pe}$ maximizing the $B_z$-spectrum from the simulation (black solid line) and
from Eq. \eqref{eq:ksat1} (black dashed line).
Temporal evolutions of (d) $v_i/c$, (e) $a_i$ and (f) $T_{e,iy}/m_ec^2$ from the simulation (black solid lines) and from Eqs. \eqref{eq:qlvi2}, \eqref{eq:qlani1}
and \eqref{eq:qltyi3}, respectively (black dashed lines), extracting $S_p$ from the simulation [black solid line of (a)]. 
The analytical predictions, Eqs. \eqref{eq:lambda_analytic}, \eqref{eq:anifinal}-\eqref{eq:vifinal}, are superposed as red dashed lines.
}
\end{figure*}

Let us confront these results to our quasilinear model. Its validity is first verified by solving the dispersion relation with the plasma parameters
measured at the saturation time. We obtain $\vert \xi_\mathrm{sat} \vert \simeq 0.5<1$, in good agreement with the approximate value of $0.4$
obtained in Sec. \ref{subsec:dominant_mode}.  This should be contrasted with the value $\xi_\mathrm{sat} \simeq 1.2 > 1$ associated with the
initial state of the system.  The dashed black lines in \mbox{Figs. \ref{fig:vi02_ve0_m100}(d,e,f)} plot $v_i(t)$, $a_i(t)$ and $T_{iy}(t)$ as
predicted by Eqs. \eqref{eq:qlvi2}, \eqref{eq:qlani1} and \eqref{eq:qltyi3} using $S_p(t)$ from the simulation [\mbox{Fig. \ref{fig:vi02_ve0_m100}(b)}].
Satisfactory agreement with the PIC results is found for the three curves. The weak variations of $v_i$ are well reproduced over the whole simulation
time. More interestingly, the pronounced variations (by about an order of magnitude) of $a_i$ and $T_{iy}$ are quantitatively captured up to
$t\omega_{pe}\simeq 1400$. Later on, our model underestimates by a factor of $\sim2$ the increase in $T_{iy}$, and consequently, overestimates $a_i$
by  the same factor. The accuracy of Eq. \eqref{eq:ksat1}, giving $k_\mathrm{sat}$ as a function of $a_i$ (here taken from the PIC curve in
\mbox{Fig. \ref{fig:vi02_ve0_m100}(e))}, is illustrated by the dashed black line in Fig. \ref{fig:vi02_ve0_m100}(c). Good agreement is observed
between the PIC and approximate curves of $k_\mathrm{sat}(t)$. 
  
Let us now return to the reference simulation of $v_e=v_i=\pm 0.2c$, the other parameters being identical to those of the previous case. The corresponding results
are displayed in Figs. \ref{fig:vi02_ve02_m100}(a-f). The strong deceleration of each electron beam is illustrated in \ref{fig:vi02_ve02_m100}(c). At the saturation
time of the ion Weibel instability ($t\omega_{pe} \simeq 500$), the electron drift velocity, $v_e$, has decreased by more than a factor of 2 and the electron
anisotropy ratio, $a_e$, has dropped to $a_e \simeq -0.2 $ (not shown). Using the instantaneous plasma parameters ($T_{i,ex}\simeq 0.01m_ec^2$, 
 $T_{i,ey} \simeq 0.03m_ec^2$, $v_i\simeq\pm0.2c$ and $v_e\simeq\pm 0.08c$), the dispersion relation gives $\vert\xi_e\vert \simeq 0.04 \ll 1$ and $\vert\xi_i\vert \simeq 0.4$, the latter value
closely matching the theoretical expectation. Plugging the simulated $S_p(t)$ values into Eqs. \eqref{eq:qlvi2}, \eqref{eq:qlani1} and \eqref{eq:qltyi3} yields
approximate curves (dashed black lines) that reproduce the PIC curves (black solid lines) to within a factor of $\sim 2$ (for $a_i$ and $T_{iy}$). Furthermore,
the $k_\mathrm{sat}$ estimate, Eq. \eqref{eq:ksat1}, is seen to underestimate the simulation values by 40\% for $t\omega_{pe}\gtrsim 500$ [Fig. \ref{fig:vi02_ve02_m100}(b)].

\begin{figure}[tbh!]
\centerline{\includegraphics[scale=1]{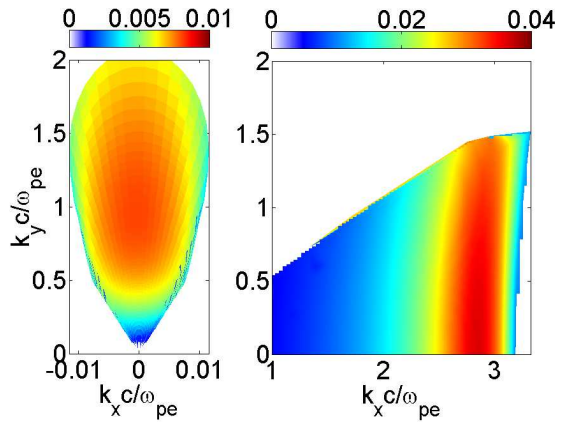}}
\caption{\label{fig:gamma2d_vi4e-1}
Growth rate $\Gamma(k_x,k_y)/\omega_{pe}$ of a colliding-beam system with $v_i= \pm 0.4c$, $v_e=0$, $T_{e,i}/m_ec^2 = 0.01$ and $m_i/m_e=1836$.
The real frequency vanishes for the unstable modes around $k_x=0$ (left panel), while the fastest-growing modes around $(k_x,k_y) \simeq (0.3,0)\omega_{pe}/c$
(right panel) propagate at a phase velocity of $\simeq 0.4c$.
}
\end{figure}

Figures \ref{fig:vi04_ve0_m1836}(a-f) show the results of a simulation run with a realistic proton mass, $m_i=1836m_e$, and $v_i=\pm 0.4c$ and $v_e=0$.
The other parameters are kept unchanged. A cruder discretization was employed for this simulation ($\Delta x=\Delta y=0.2c/\omega_{pe}$). The theoretical
growth rate map shown in \mbox{Fig. \ref{fig:gamma2d_vi4e-1}} predicts that the system is initially dominated by a longitudinal electrostatic mode of growth rate
$\Gamma_\mathrm{max}/\omega_{pe} \simeq 0.04$, wave number $k_xc/\omega_{pe} \simeq 2.8$ and phase velocity $\omega/k \simeq \pm 0.4c$. This is 
indicative of a Buneman instability, driven by a relative electron-ion velocity exceeding the electron thermal velocity ($v_{te}=0.1c$). These predictions
account for the initially dominant $E_x$ energy observed in \mbox{Fig.  \ref{fig:vi04_ve0_m1836}(a)}. Following an exponentially growing phase (at a rate
$\Gamma/\omega_{pe} \simeq 0.03$, comparable to the theoretical value), the $E_x$ energy saturates around $t\omega_{pe}\simeq200$, and rapidly
decays away. The $B_z$ magnetic energy associated with the Weibel instability 
overcomes the
$E_x$ energy at $t\omega_{pe}\simeq400$, before saturating at $t\omega_{pe}\simeq1000$. As in the previous cases, the magnetic energy remains
approximately constant after saturation. The evolution of $k_\mathrm{sat}$ is well reproduced by Eq. \eqref{eq:ksat1} for $t\omega_{pe}>1000$
[Fig. \ref{fig:vi04_ve0_m1836}(c)].  Furthermore, the temporal evolutions of $v_i$, $a_i$ and $T_{iy}$ in the simulation are well reproduced by the
quasilinear theory over the whole simulation time [Figs.  \ref{fig:vi04_ve0_m1836}(d,e,f)]. Surprisingly good agreement is found during the early Weibel
phase, although the $\vert \xi_i\vert <1$ condition is not fulfilled at the saturation time: $\vert \xi_i\vert \simeq 1.2$ is then obtained for the measured parameters
$T_{ex} = T_{ix,y}=0.2m_ec^2$, $T_{e,iy}=0.3m_ec^2$, $v_i=\pm0.4c$, $v_e=0$.  Later on, $\vert \xi_i\vert $ steadily decreases below unity, reaching
$\vert \xi_i\vert \simeq 0.8$ at $t\omega_{pe}=4000$.

\subsubsection{3-D periodic simulations} \label{subsubsec:pic3d}

\begin{figure*}
\centerline{\includegraphics[scale=1]{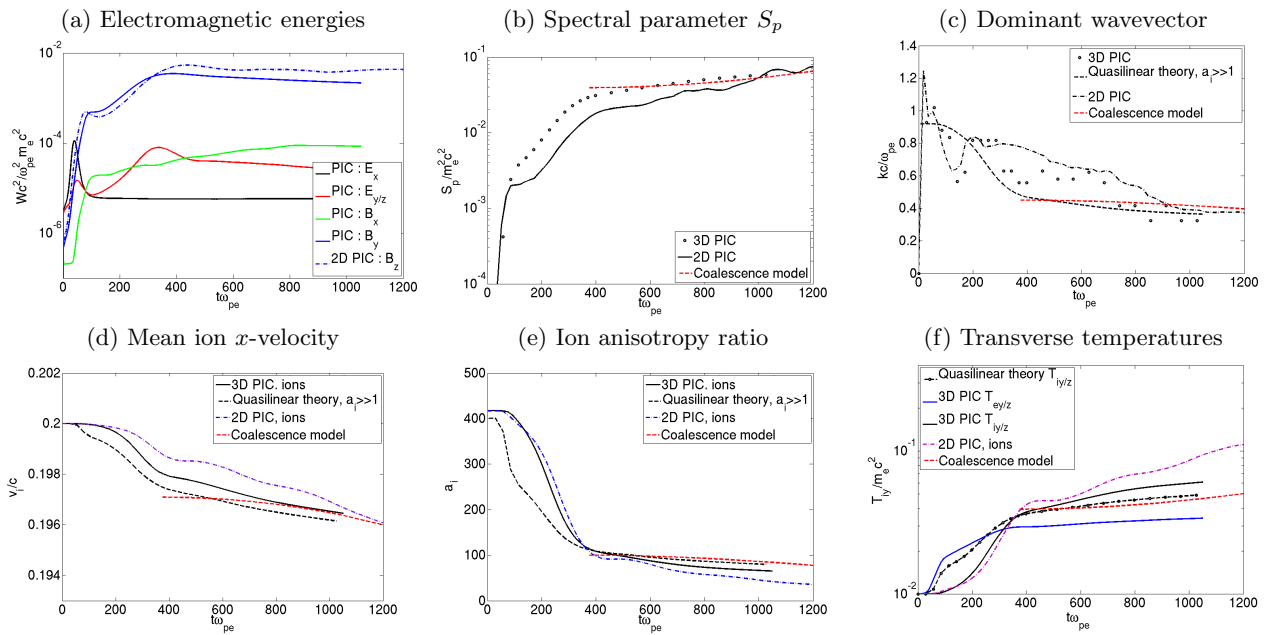}}
\caption{\label{fig:vi02_ve02_m100_3d}
3-D simulation with initial conditions: $v_i =\pm0.2c$, $v_e=0$, $T_{e,i}/m_ec^2=0.01$, $m_i/m_e=100$.
(a) Temporal evolution of the mean electromagnetic energies (normalized to $m_ec^2\omega_{pe}^3/c^3$). We have checked that the $B_y$ and $B_z$
histories exactly coincide. The $B_z$ history from the 2-D simulation is superposed as a blue dotted-dashed line.
(b) Temporal evolution of $S_p/(m_ec)^2$, defined by Eq. \eqref{eq:sp} from the simulation (circles). 
(c) Temporal evolution of $k_\mathrm{sat}c/\omega_{pe}$ maximizing the $B_z$-spectrum from the simulation (circles) and from Eq. \eqref{eq:ksat1} (red dashed line).
Temporal evolutions of (d) $v_i/c$, (e) $a_i$ and (f) $T_{e,iy}/m_ec^2$ from the simulation (solid lines) and of Eqs. \eqref{eq:qlvi2}, \eqref{eq:qlani1} and \eqref{eq:qltyi3},
respectively (black dashed lines) measuring $S_p$ from the simulation (circles of (a)). 
The 2-D PIC simulation results of Fig. \ref{fig:vi02_ve02_m100} are superposed as black dotted-dashed lines.
The analytical predictions of Eqs. \eqref{eq:lambda_analytic}, \eqref{eq:anifinal}-\eqref{eq:vifinal} are superposed as red dashed lines.}
\end{figure*}

The above quasilinear equations can be readily generalized to 3-D systems, where the Weibel instability develops in the $y-z$ transverse plane, given
the following definition of $S_p$:
\begin{equation}
  S_p^{3D} = e^2 \sum_{k_y,k_z} \frac{\langle \vert DFT_{y,z}(B_y) \vert^2 +   \vert DFT_{y,z}(B_z ) \vert^2 \rangle_x }{k_y^2 + k_z^2}  \, . \label{eq:sp3d}
\end{equation}
Moreover, the dispersion relation, Eq. \eqref{eq:dispew1}, remains unchanged when shifting from 2-D to 3-D non-relativistic bi-Maxwellians. 

Figures \ref{fig:vi02_ve02_m100_3d}(a-f) present the results of a 3-D periodic simulation using the same plasma parameters as in Figs. \ref{fig:vi02_ve02_m100_3d}(a-f):
$v_e=v_i=\pm 0.2c$, $T_{e,i}=0.01m_ec^2$ and $m_i=100m_e$. The simulation domain has dimensions $102.4 c/\omega_{pe} \times 96 c/\omega_{pe} \times 96 c/\omega_{pe}$
with the discretization $\Delta x = \Delta y = \Delta z = 0.2c/\omega_{pe}$. Each cell is initialized with 30 macro-particles per species.

The 2-D and 3-D simulations give very similar results with respect to both the electromagnetic and kinetic quantities (Figs. \ref{fig:vi02_ve02_m100} and
\ref{fig:vi02_ve02_m100_3d}). A somewhat surprising finding is that the 2-D simulation predicts a slightly faster increase in the transverse ion temperature,$T_{iy}$.  
The observed overall agreement between 2-D and 3-D simulations  is consistent with Ref. \cite[]{Silva_2006},
where it was demonstrated that the multidimensional physics of unstable two-stream systems is well captured by 2-D simulations resolving the drift (longitudinal)
direction. Finally, as in the 2-D case, the predictions from quasilinear theory reasonably match the 3-D simulation results during the ion Weibel-saturation stage
($t\omega_{pe}\gtrsim 400$).

\subsection{Coalescence-driven spectral dynamics}

From the quasilinear theory of the transverse Weibel instability, we have derived simple analytical relations between the ion parameters and the spectral
quantity $S_p=e^2\sum_k \vert A_k \vert^2$. These equations have been shown to match the PIC simulation results for various plasma parameters,
provided the ion anisotropy remains large enough in the nonlinear stage. A closure relation relation, however, must be provided to get a fully predictive
model.

Previous studies of the Weibel instability revealed that the nonlinear filament dynamics is subject to secondary processes, such as kink instabilities 
\cite[]{apj_Mil_Nakar_2006}, which generate $k_x\neq 0$ modes in the magnetic spectrum, or filament coalescence \cite[]{Honda_2000, Medvedev_2005,Polomarov_2008,Gedalin_2010}.
The latter mechanism originates from the partial neutralization by the background electrons of the ion current filaments formed in the nonlinear stage. This results
in a nonvanishing magnetic attraction between neighboring filaments of same current sign, which then tends to coalesce. Each merger generates a larger filament,
of roughly twice the size of the primary filaments,  thus leading to increasingly low-$k_y$ modes in the magnetic spectrum. The close agreement between the above
simulations and the quasilinear theory of the purely transverse ion Weibel instability indicates that the long-term dynamics of the latter is mostly governed by the
evolution of the $k_x=0$ modes, and therefore by coalescence effects.  In the framework of our model, this  proceeds along the following lines. As the ions get heated,
the dominant wave vector, $k_\mathrm{sat}$, decreases accordingly to Eq. \eqref{eq:ksat1}. Since,  according to our simulations, $\langle B_z^2\rangle$ remains
essentially constant at late times, there follows a steady increase in $S_p \sim \langle B_z^2\rangle /k_\mathrm{sat}^2$, and consequently of the ion heating.
This picture should hold as long as the ion anisotropy is sufficient to sustain the instability and/or the filament size remains below the transverse size of the system. 
We now propose to derive, from simple coalescence arguments \cite[]{Medvedev_2005, Achterberg_2007b}, a closure equation for our model describing the
temporal evolution of $k_\mathrm{sat}$.

\subsubsection{Collective dynamics of the current filaments} \label{subsubsec:coal}

\begin{figure}
\centerline{\includegraphics[scale=1]{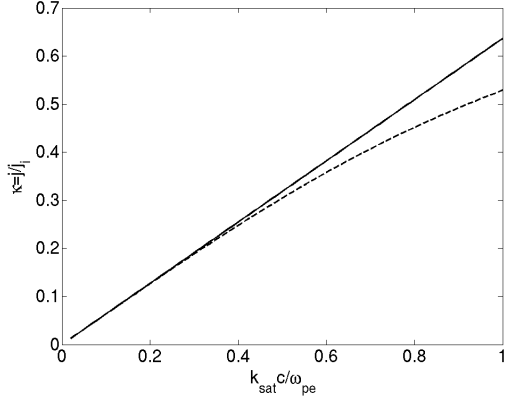}}
\caption{\label{fig:coal_medvedev}
Screening factor Eq. \eqref{eq:ecrantage} as a function of $k_\mathrm{sat}c/\omega_{pe}$ (black solid line). The estimate $j_x/j_{ix}=2ck_\mathrm{sat}/\pi\omega_{pe}$
is superposed as a black dashed line.}
\end{figure} 

A critical parameter ruling the current filament dynamics is the average net current, $j_x$, carried by the electron-ion filaments \cite[]{Achterberg_2007b}. The ion current
contribution, $j_{ix}$, can be estimated assuming spatial separation of the (initially superposed) counter-streaming ion beams: $j_{ix} \simeq Z_i e n_i v_i/2$, where
$n_i$ is the total density of the ion population (including the two beams). For the reference simulation ($v_i=v_e=\pm 0.2c$, $m_i=100m_e$), one thus
predicts $j_{ix} \simeq 0.1en_cc$, in agreement with the current profile of  \mbox{Fig. \ref{fig:filam_ion}(d)}. The electrons tend to neutralize the ion current with
an efficiency increasing with the size of the filament relative to the electron skin depth. Assuming a linear, non-relativistic electron response,  the screening factor,
$\kappa=\vert j_x/j_{ix}\vert$, of a cylindrical ion current filament of diameter $\lambda_\mathrm{sat}/2=\pi/k_\mathrm{sat}$, reads \cite{Achterberg_2007b}
\begin{equation}\label{eq:ecrantage}
  \kappa = 2 I_1\left(\frac{\pi\omega_{pe}}{2ck_\mathrm{sat}} \right) K_1\left(\frac{\pi\omega_{pe}}{2ck_\mathrm{sat}}\right) \, ,
\end{equation}
where $I_1$ and $K_1$ are the modified Bessel functions of the first and second kind, respectively. In the limit of $\pi\omega_{pe}/2ck_\mathrm{sat} \gg 1$, the
above equation simplifies to
\begin{equation} \label{eq:ecrantage_app}
  \kappa \simeq  j_{i_x} \frac{2ck_\mathrm{sat}}{\pi\omega_{pe}}  \, .
\end{equation}
It can be shown that \mbox{Eq. \eqref{eq:ecrantage_app}} also holds in a planar geometry. Both expressions are plotted
in Fig. \ref{fig:coal_medvedev}.  Applied to the current profiles of Fig. \ref{fig:filam_ion}(d), exhibiting a typical wavelength of $\lambda_\mathrm{sat}\omega_{pe}/c\simeq10$,
the above formula predicts  a screening electron current of  $\vert j_{ex} \vert \simeq 0.5j_{ix}$, matching the simulation results. From \mbox{Eq. \eqref{eq:ecrantage_app}},
we deduce the approximate net filament current
\begin{equation}\label{eq:screened_current}
  j_x \simeq Z_i e n_i v_i \frac{ck_\mathrm{sat}}{\pi\omega_{pe}}\, .
\end{equation}

We now derive a differential equation obeyed by the average separation length of the filaments, which will be be equated to $\lambda_\mathrm{sat}$. 
Furthermore, the diameter of the filaments will be approximated to $\lambda_\mathrm{sat}/2$, while their mean particle and current density will be assumed constant
during each merging process.  Although similar, our approach differs from that of Refs. \cite{Medvedev_2005, Achterberg_2007b}. The equation of motion applied
to the distance between two filaments, $Y$, reads
\begin{equation}\label{eq:pfd1}
  \ddot{Y} \simeq  -\frac{j_x}{m_i n_i} \langle B_z \rangle  \simeq - \frac{Z_i e \kappa v_i}{m_i} \langle B_z \rangle  \, ,
\end{equation} 
where $\langle B_z \rangle $ denotes the averaged magnetic seen by a filament, which derives from the averaged vector potential $\langle A_x \rangle$.
The above equation can then be recast as
\begin{equation}\label{eq:pfd2}
  \Delta \dot{Y}^2   \simeq  \frac{2Z_i e \kappa v_i}{m_i} \Delta  \langle A_x \rangle (t) \, ,
\end{equation}
Making the approximations $e\langle A_x \rangle \simeq S_p^{1/2}$ and $e\Delta \langle A_x \rangle \simeq S_p^{-1/2}\Delta S_p /2$ leads to
\begin{equation}\label{eq:pfd3}
  \dot{Y}^2  \simeq Z_i \int_{S_p(t^*)}^{S_p(t)} \frac{\kappa  v_i}{2\sqrt{S_p}}  d S_p \, ,
\end{equation} 
where $t_*$ denotes the start time of the nonlinear phase and we have assumed $\dot{Y}(t_*)=0$. The ions are assumed to fulfill $\vert \xi_i(t_*) \vert \ll 1$.
Let us now introduce $\tau_c$, the typical coalescence time between two filaments, such that $\vert \dot{Y} \vert \sim  \lambda_\mathrm{sat}/\tau_c$
\cite{Medvedev_2005, Achterberg_2007b}. During a merging event, $\vert \dot {\lambda}_\mathrm{sat}\vert  \sim \lambda_\mathrm{sat}/\tau_c$, so that
we can estimate $\vert \dot {\lambda}_\mathrm{sat}\vert  \sim \vert \dot{Y} \vert$. There follows
\begin{equation}\label{eq:pfd4}
  \dot{\lambda}_\mathrm{sat}^2 \simeq Z_i \int_{S_p(t_*)}^{S_p(t)} \frac{\kappa  v_i}{2\sqrt{u}}  d u \, .
\end{equation}
where we have further assumed $\dot{\lambda}_\mathrm{sat}(t_*) = 0$. Since this equation only involves quantities spatially averaged over a large
number of filaments, it can be combined to the quasilinear equations of Sec. \ref{sec:tepp}. 

To make analytical progress, we inject in Eq. \eqref{eq:pfd4} simplified forms of Eqs. \eqref{eq:qlvi}, \eqref{eq:qltyi3}, \eqref{eq:qlani1}, \eqref{eq:qlsp}
and \eqref{eq:ecrantage}, in the limits of $K_2 \simeq m_iv_0^2$ and $a_i \gg 1$ : 
\begin{align}
S_p &\simeq \frac{1}{2\alpha_i  Z_i^2 }\left( \frac{m_iv_0}{2+a_{i}}  \right)^2\, , \label{eq:spi2} \\
k_\mathrm{sat} &\simeq  \frac{\omega_{pi}}{2c}  \left( \frac{m_iv_0 }{\sqrt{2 \alpha_i Z_i^2S_p}   } \right)^{1/2} \label{eq:ksat3}\, , \\
a_i &\simeq  \frac{m_iv_0 }{\sqrt{2 \alpha_i Z_i^2S_p}  } \, , \label{eq:ani2} \\
T_{iy} &\simeq v_0 \sqrt{2 \alpha_i  Z_i^2S_p} \, , \label{eq:tiy2} \\
v_i &\simeq  v_0 \exp\left[-\frac{2\sqrt{2 \alpha_i Z_i^2 S_p}}{m_iv_0 }\right] \, , \label{eq:vi2} \\
\kappa &\simeq  \frac{2ck_\mathrm{sat}}{\pi\omega_{pe}} \, . \label{eq:ecrantage2}
\end{align}
Plugging Eqs. \eqref{eq:spi2}-\eqref{eq:ecrantage2} into Eq. \eqref{eq:pfd4} with $v_i=v_0$ yields 
\begin{equation} \label{eq:coalint1}
  \dot{\lambda}_\mathrm{sat}^2 \simeq  \frac{\omega_{pi} v_0}{ 2\pi m_i\omega_{pe}}  \int_{S_p(t_*)}^{S_p(t)} \frac{du}{\sqrt{u}} \left( \frac{m_iv_0 }{\sqrt{2 \alpha_iZ_i^2 u}}\right)^{1/2}\, .
\end{equation}
The above equation can be readily solved in combination with Eq. \eqref{eq:ksat3}, giving
\begin{equation}
  \lambda_\mathrm{sat} \simeq \lambda_* \left(1+ \frac{\Delta t^2}{\tau_0^2}  \right)\, ,\label{eq:lambda_analytic} \\
  \end{equation}
where $\Delta t \equiv t-t_*$ and
\begin{equation}\label{eq:tau0}
  \tau_0 = \frac{2\pi (8\alpha_i)^{1/4}} {v_0} \left(\frac{m_i}{Z_im_e}\right)^{1/4} \sqrt{\frac{\lambda_*c}{\omega_{pi}}}  \, 
\end{equation}
is the typical coalescence time, that is, the time over which $k_\mathrm{sat}$ decreases by half. It can also be viewed as the lifetime of the slowly-evolving
filamentary state established at magnetic saturation. As expected, it increases with the typical distance between filaments after saturation, $\lambda_*$. Note that
the dependence of $\tau_0$ upon $\lambda_*$ cancels out in the long-time limit of $\lambda_\mathrm{sat}(t)$. Also, the influence of the electron screening transpires
through the $m_e^{-1/4}$ term. 

The last step consists in substituting Eq. \eqref{eq:lambda_analytic} into \eqref{eq:spi2}-\eqref{eq:vi2} to obtain a fully predictive analytical formulation of the plasma
parameters as a function of $\Delta t$ and of the wavevector at the end of the linear phase ($k_*$):
\begin{align}
a_i &\simeq  \frac{4k_*^2c^2}{\omega_{pi}^2} \frac{1}{(1 + \Delta t^2/\tau_0^2)^2} \, , \label{eq:anifinal} \\
S_p &\simeq \frac{m_i^2v_0^2}{2\alpha_iZ_i^2 } \frac{(1 + \Delta t^2/\tau_0^2)^4}{\left(4k_*^2c^2/\omega_{pi}^2 + 2(1 + \Delta t^2/\tau_0^2)^2  \right)^2}\, , \label{eq:spifinal} \\
T_{iy} &\simeq m_iv_0^2  \frac{(1 + \Delta t^2/\tau_0^2)^2}{4k_*^2c^2/\omega_{pi}^2 + 2(1 + \Delta t^2/\tau_0^2)^2  } \, , \label{eq:tiyfinal} \\
v_i &\simeq v_0\exp \left[-  \frac{2(1 +\Delta t^2/\tau_0^2)^2}{4k_*^2c^2/\omega_{pi}^2 + 2(1 + \Delta t^2/\tau_0^2)^2  }\right] \, . \label{eq:vifinal} 
\end{align} 
Taking $\Delta t = 0$ gives the plasma parameters at the beginning of the Weibel saturation phase ($k_\mathrm{sat}=k_*$).

The cold-limit approximation made in deriving Eqs. \eqref{eq:spi2}-\eqref{eq:ecrantage2} (\emph{i.e.}, assuming  $T_{i,e}(0)=0$) is valid provided the initial
temperature verifies
\begin{equation}\label{condition_model}
T_{i,x,y}(0) \ll T_{iy}(t_*)  \simeq \frac{m_iv_0^2   }{4k_*^2c^2/\omega_{pi}^2  +2 } \, ,
\end{equation}
a condition fulfilled in the previous simulations.

\subsubsection{Influence of the initial filament size}\label{subsubsec:initcond}

Our model requires the knowledge of the typical filament wavelength at the beginning of the nonlinear ion-Weibel phase, $\lambda_*$.
A crude approximation of $\lambda_*$ can be made using Davidson's magnetic trapping model \cite{Davidson_1972}.
This model assumes that the linear phase of the instability ceases when the magnetic bounce frequency of the driving particles (here the ions)
becomes comparable to the linear growth rate of the instability. The saturated potential vector, $A_*\equiv A(t_*)$, is therefore expected to fulfil
\begin{equation}
  \Gamma_{k_*} \simeq \sqrt{\frac{Z_i e v_0 k_*^2 A_*}{m_i}} \, ,
\end{equation}
which leads to
\begin{equation}\label{eq:iondavion2}
  A_* \simeq \frac{m_i\Gamma^2_{k_*}}{Z_i e v_0 k_*^2}\, .
\end{equation}
Making use of $\xi_i = \Gamma/k_y\sqrt{2T_{i0}/m_i}$ and $S_p(t_*) \sim (eA_*)^2 $, we obtain
\begin{equation}\label{eq:spo}
  S_p(t_*) \simeq \frac{T_{i0}^2}{Z_i^2v_0^2 } \xi_i^{4} \, .  
\end{equation} 
Combining Eqs. \eqref{eq:ksat1} and \eqref{eq:qlani1} allows us to estimate $k_*$ as 
\begin{equation}
  k_* \simeq \frac{\omega_{pi}}{2c} \sqrt{   \frac{K_2}{T_{i0} \sqrt{1+\frac{2\alpha_iK_2}{m_iv_0^2}\xi_i^4 }}   -2}\, . \label{eq:ket}
\end{equation}
The $\xi_i$ term can be estimated by maximizing the growth rate computed from the exact dispersion relation (using the initial plasma parameters).
For initially low-temperature plasmas, the initial ion anisotropy ratio verifies $a_i^{(0)} \sim m_i v_0^2/T_i^{(0)} \gg 1$, so that $\lambda_* \ll  c/\omega_{pi}$.
The above formulation, based on the Davidson scaling, depends on $\xi_i^4(0)$ and is thus imprecise. For $v_i =\pm 0.2c$, $v_e=0$
and $m_i=100m_e$, linear theory predicts $\xi_i\simeq 3.8$, and hence $\lambda_*\omega_{pe}/c\simeq 17$ 
(to be compared
with the simulation result $\lambda_*\omega_{pe}/c  \simeq 12.6$ at $\omega_{pe}t_*=700$) and $\tau_0\simeq 4300\omega_{pe}$. 
For $v_i=\pm 0.4c$ and $m_i = 1836m_e$, one obtains $\xi_i\simeq 1.8$, $\lambda_*\omega_{pe}/c\simeq 9.3$ 
with the simulation result $\lambda_*\omega_{pe}/c  \simeq 14$ at $\omega_{pe}t_*=1000$) and  $\tau_0\simeq 1500\omega_{pe}$. Yet, the error made in
using these estimates should not impact the long-term evolution of $\lambda_\mathrm{sat}$: 
\begin{equation}\label{eq:evolongterm}
  \lambda_\mathrm{sat}(\Delta t \gtrsim 3\tau_0) \simeq  \frac{1}{12\pi^2(2\alpha_i)^\frac{1}{2}} \left(\frac{v_0}{c}\right)^2  \frac{c}{\omega_{pe}}  (\omega_{pi}\Delta t)^2 \, ,
\end{equation}
which is independent of $\lambda_*$, yet involves the electron mass. Using Eq. \eqref{eq:ksat1}, we can derive the long-time expression of the ion anisotropy ratio: 
\begin{align}
  a_i 
\simeq 2048 \pi^6 \alpha_i \frac{m_i}{Z_im_e}  \left(  \frac{c}{v_0} \right)^4  (\omega_{pi}\Delta t)^{-4} \, .\label{eq:evolongterm_ani}
\end{align}
The time required to reach quasi-isotropization ($a_i=2$) can therefore be estimated as 
\begin{equation}\label{eq:dtmax_ani}
\Delta t_\mathrm{form} \simeq \frac{26}{\omega_{pi}}  \frac{c}{v_0} \left(\frac{Z_i m_i}{m_e}\right)^{1/4} \, ,
\end{equation}
where we have assumed $a_i^{(0)} \gg 1$. This time can be viewed as a lower limit of the shock formation time since, in addition to neglecting the initial electron-driven phase,
our calculation stops before full isotropization ($a_i=0$).  Interestingly, this lower limit scales as $m_i^{3/4}/m_e^{1/4}$, as a result of electron screening. Our formulae will
be compared with shock simulations in a forthcoming publication.

\begin{figure}
\centerline{\includegraphics[scale=1]{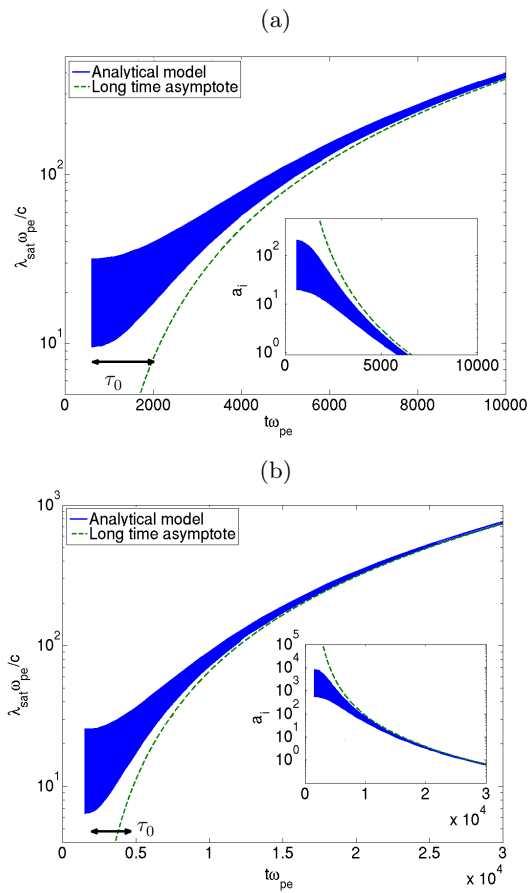}}
\caption{\label{fig:errorinit}  
Temporal evolution of $\lambda_\mathrm{sat}\omega_{pe}/c$ from the numerical (red solid line) and analytical approximate (blue solid line) resolution
of Eq. \eqref{eq:coalint1} for two systems:
(a) $m_i=100m_e$, $v_i=\pm0.2c$ with the initial conditions $ 9\le \lambda_*\omega_{pe}/c \le 30$ at $t_*\omega_{pe}=600$;
(b) $m_i=1836m_e$, $v_i=\pm0.4c$ with the initial conditions $6\le \lambda_*\omega_{pe}/c \le 25$ at $t_*\omega_{pe}=1500$.
The subpanels plot the temporal evolution of the ion anisotropy ratio $a_i$, the typical time $\tau_0$, Eq. \eqref{eq:tau0}, being indicated by a black arrow.
The long-time approximations of Eqs. \eqref{eq:evolongterm} and \eqref{eq:evolongterm_ani} are superposed as green dashed lines. 
}     
\end{figure}

Figures \ref{fig:errorinit}(a,b) illustrate for two parameter sets the theoretical  evolution of $\lambda_\mathrm{sat}$ and $a_i$. In each case, we have considered a
finite range of values for $\lambda_*$. As expected in the $a_i \gg 1$ limit, the curves converge to the same limiting curve after a few $\tau_0$'s.
Finally, Eqs. \eqref{eq:anifinal}-\eqref{eq:vifinal} are plotted in Figs. \ref{fig:vi02_ve0_m100}-\ref{fig:vi04_ve0_m1836} and \ref{fig:vi02_ve02_m100_3d},
where they show overall agreement with the corresponding simulation results.

\section{Conclusions}\label{sec:conclusions}

We have described the slow dynamics of the non-linear Weibel-filamentation instability using a set of simplified quasilinear relations, valid in the case
of highly anisotropic ion beams and homogeneous profiles. Fairly good agreement between the theoretical expectations and 2-D/3-D simulations in the
non-linear regime has been found for various plasma parameters. A closure relation modeling the collective filament dynamics has then been derived
and solved in the high-ion anisotropy limit. Our analytical formulae, Eqs. \eqref{eq:anifinal}-\eqref{eq:vifinal}, are found to capture with reasonable accuracy the
simulation results.  We have obtained an expression for the ``quasi-isotropization time'' of the ion population, Eq. \eqref{eq:dtmax_ani},  which may be considered
as an upper limit of the shock formation in the case of bounded ion beams. An important finding is that this time scales as $\omega_{pi}^{-1}(m_i/m_e)^{1/4}$ 
due to electron screening effects. This result should be taken in consideration when analyzing the results of kinetic simulations run, as is usual, with nonphysical ion
\cite[]{Kato_2008} or electron masses \cite[]{Fox_2013,Huntington_2013}. Our non-linear model, albeit based upon a number of simplifying assumptions,
therefore constitutes a complementary tool to first-principles simulations for the understanding of the ion-Weibel-filamentation instability in realistic settings. 
Its applicability to shock-relevant configurations will be addressed in a forthcoming paper.

\section*{Acknowledgments}

The authors gratefully acknowledge Anne Stockem and Frederico Fiuza for interesting discussions. 
The PIC simulations were performed using HPC resources at TGCC/CCRT (Grant No. 2013-052707).

\appendix \label{sec:appendix}

\section{Electromagnetic dispersion relation for bi-Maxwellian distributions}

Let us consider a charge-neutral plasma composed of a number of charged particle species (identified by the $s$ subscript). The linearized Vlasov-Maxwell
equations yield the general electromagnetic dispersion relation of the plasma between the wave vector, $\mathbf{k}$, and the imaginary frequency, $\omega$
\cite[]{Ichimaru_1973}:
\begin{align}
  &(\omega^2 \varepsilon_{xx}-k^2c^2 \sin^2\theta)(\omega^2 \varepsilon_{yy}-k^2c^2 \cos^2\theta) \nonumber\\
  &-(\omega^2 \varepsilon_{xy}+k^2c^2 \cos \theta \sin\theta )^2=0\, . \label{eq:dispe_2d}
\end{align}
where $\theta$ denotes the angle between the $x$-axis and $\mathbf{k}$. The dielectric tensor can be expressed in the form
\begin{equation}
\boldsymbol{\epsilon}_{\alpha\beta} = \delta_{\alpha\beta}+ \sum_s \frac{\omega_{ps}^2}{\omega^2} \boldsymbol{\chi}_s \, ,
\end{equation}
where $\chi_s$ is the susceptibility tensor and $\omega_{ps}$ is the plasma frequency of the $s$th charged species. In the case of non-relativistic bi-Maxwellian
distributions [Eq. \eqref{eq:bm}], the tensor elements read
\begin{widetext}
\begin{align}
\chi_{xx}=&  -1 +  \left[\cos^2 \theta + \sin^2 \theta \frac{\mu^2}{\mu_\perp^2} - \sin 2\theta \frac{\mu}{\mu_\perp} \right ]\left[ 1 -\xi^2 \mathcal{Z}'(\xi)\right]
+ 2 \sin \theta \sqrt{\frac{\mu_{\parallel}}{2}} \frac{\mu}{\mu_\perp} \beta_0  \mathcal{Z}(\xi)   \nonumber\\
&-   \left[ \beta_0^2\mu_{\parallel} +\frac{\mu_{\parallel}}{\mu_\perp}\sin^2 \theta -2 \sin^2\theta   \frac{\mu^2}{\mu_\perp^2} +2\sin 2\theta \frac{\mu}{\mu_\perp} \right] \frac{\mathcal{Z}'(\xi)}{2} \nonumber \\
&-2  \sqrt{\frac{\mu_{\parallel}}{2}} \beta_0 \left[\cos \theta -\sin \theta   \frac{\mu}{\mu_\perp}  \right] \xi\mathcal{Z}'(\xi) \, , \label{eq:xibm1} \\
\chi_{yy}=& - 1 -\left[ \cos^2 \theta  \frac{\mu_{\parallel}}{\mu_\perp} -2 \cos^2 \theta \frac{\mu^2}{\mu_\perp^2} 
-2\sin 2\theta \frac{\mu}{\mu_\perp} \right]\frac{\mathcal{Z}'(\xi)}{2}  \nonumber\\    
&+ \left[\sin^2 \theta + \cos^2 \theta \frac{\mu^2}{\mu_\perp^2} + \sin 2\theta \frac{\mu}{\mu_\perp} \right]
\left[ 1 - \xi^2 \mathcal{Z}'(\xi)\right]  \, , \label{eq:xibm2} \\
\chi_{xy}=& - \left[\cos \theta \sin \theta (1- \frac{\mu^2}{\mu_\perp^2})+\cos 2\theta \frac{\mu}{\mu_\perp}\right] \xi 
\left[\mathcal{Z}(\xi) +\xi \mathcal{Z}'(\xi) \right]
- \frac{\mu}{\mu_\perp} \sqrt{\frac{\mu_{\parallel}}{2}} \beta_0 \cos \theta \mathcal{Z}(\xi)\nonumber\\
&- \cos \theta \sin \theta  \left[ \frac{\mu_{\parallel}}{\mu_\perp}-1- \frac{\mu^2}{\mu_\perp^2} \right] \frac{\mathcal{Z}'(\xi)}{2} 
-\left[\sin \theta +\cos \theta \frac{\mu}{\mu_\perp} \right] \beta_0 \sqrt{\frac{\mu_{\parallel}}{2}}\xi\mathcal{Z}'(\xi) \, , \label{eq:xibm3}
\end{align}
\end{widetext}
where $\mathcal{Z}$ and $\mathcal{Z'}$ are the plasma dispersion function and its derivative. We have also introduced $\beta_0=v_0/c$ and
\begin{align}
\mu=&\cos\theta \sin \theta \left(\frac{m}{T_x} - \frac{m}{T_y}\right)   \, ,  \\
\mu_{\parallel}=& \frac{m \cos^2 \theta}{T_x} + \frac{m\sin^2 \theta}{T_y}  \, , \\
\mu_\perp =& \frac{m \sin^2\theta}{T_x} + \frac{m\cos^2 \theta}{T_y} \, ,  \\
\xi =& \sqrt{\frac{\mu_{\parallel}}{2}} \left(\frac{\omega}{k} - v_0\cos \theta  \right)  \, . \label{eq:X}
\end{align}

Introducing the normalized wave phase velocity, $\beta_\phi=\omega/kc$, Eq. \eqref{eq:dispe_2d} can be recast as
\begin{equation}
  ak^4+bk^2+c=0\, , \label{eq:dispe_oblique}
\end{equation}
with
\begin{equation}
a=(\beta_\phi^2 -\sin^2\theta)(\beta_\phi^2 -\cos^2\theta) -\cos^2\theta \sin^2\theta \, , \label{eq:dispe_obliquea}\\
\end{equation}
\begin{align}
b&=(\sin^2\theta-\beta_\phi^2) \sum_s \omega_{ps}^2\chi_{yy}  (\cos^2\theta-\beta_\phi^2)\sum_s \omega_{ps}^2\chi_{zz} \nonumber\\
&+2\cos\theta \sin\theta\sum_s \omega_{ps}^2\chi_{yz}\, ,   \label{eq:dispe_obliqueb}
\end{align}
\begin{align}
c&= \Big(\sum_s \omega_{ps}^2\chi_{yy}\Big) \Big(\sum_s \omega_{ps}^2\chi_{zz} \Big) - \Big(\sum_s\omega_{ps}^2 \chi_{yz}\Big)^2\, .\label{eq:dispe_obliquec}
\end{align}
In Eqs. \eqref{eq:xibm1}-\eqref{eq:dispe_obliquec}, the subscript $s$ have been omitted on the elements of $\chi_s$ for the sake of clarity. 
The wave vector is then given by
\begin{align}
  k^2 &=\frac{-b(\beta_{\phi}) \pm \sqrt{\Delta(\beta_{\phi}})}{2a(\beta_{\phi})}\label{eq:solg1} \, ,
\end{align}
with $\Delta =\sqrt{b^2 -4ac} $.
This formulation, in which the squared wave number $k^2 (>0)$ is a function of $\beta_\phi$ only (for a given propagation angle $\theta$), 
lends itself to the efficient numerical scheme introduced by Fried and Gould \cite{Fried_Gould_1961} in a non-relativistic electrostatic framework, and generalized recently
to the electromagnetic regime for various distribution functions \cite[]{Ruyer_Gremillet_2013,Ruyer_Gremillet_2014}. This scheme consists, first, in determining the locus
of the zeroes of $\Im \mathcal{G}(\beta_\phi)$. This can be readily performed by means of a contour plot in a finely discretized portion of the complex $\beta_\phi$ plane. 
Then, we retain those zeroes fulfilling $\Re \mathcal{G}(\beta_\phi) > 0$ and identify $k=\sqrt{\Re \mathcal{G}(\beta_\phi)}$. Depending on the $\beta_\phi$-domain considered,
this method allows us to simultaneously solve for a set of discrete electromagnetic solutions $\omega(k,\theta)$.

\end{document}